\documentclass[twocolumn,aps,prb,superscriptaddress, floatfix]{revtex4-2}
\usepackage{graphicx}
\usepackage{braket}
\usepackage{hyperref}
\usepackage{xcolor, soul}

\newcommand{\Ham}{{\mathcal{H}}}
\newcommand{\Scal}{{\mathcal{S}}}
\newcommand{\tr}{{\text{Tr}}}

\begin{document}

\title{Universal shape-dependence of quantum entanglement in disordered magnets}
\author{Natalie Love}
\affiliation{Department of Physics and Astronomy, Northwestern University, Evanston, IL 60208}
\author{Istv\'{a}n A. Kov\'{a}cs}
\email{istvan.kovacs@northwestern.edu}
\affiliation{Department of Physics and Astronomy, Northwestern University, Evanston, IL 60208}
\affiliation{Northwestern Institute on Complex Systems, Northwestern University, Evanston, IL 60208}
\affiliation{Department of Engineering Sciences and Applied Mathematics, Northwestern University, Evanston, IL 60208}
\date{\today}

\begin{abstract}
Disordered quantum magnets are not only experimentally relevant, but offer efficient computational methodologies to calculate the low energy states as well as various measures of quantum correlations. We present a systematic analysis of quantum entanglement in the paradigmatic random transverse-field Ising model in two dimensions, using an efficient implementation of the asymptotically exact strong disorder renormalization group method. The phase diagram is known to be governed by three distinct infinitely disordered fixed points (IDFPs) that we study here.
For square subsystems, it has been recently established that quantum entanglement has a universal logarithmic correction due to the corners of the subsystem at all three IDFPs. This corner contribution has been proposed as an ``entanglement susceptibility'', a useful tool to locate the phase transition and measure the correlation length critical exponent. Towards a deeper understanding, we quantify how the corner contribution depends on the shape of the subsystem. 
While the corner contribution remains universal, the shape-dependence is qualitatively different in each universality class, also confirmed by line segment subsystems, a special case of skeletal entanglement. Therefore, unlike in conformally invariant systems, in general different subsystem shapes are versatile probes to unveil new universal information on phase transitions in disordered quantum systems.
\end{abstract}

\maketitle

\section*{Introduction and Motivation}

Quantum phase transitions occur in the ground state of a quantum system, by varying a quantum control parameter, resulting in complex correlation patterns.
Locating quantum phase transitions remains an open challenge in a great deal of interacting quantum systems, especially in higher dimensions. Apart from technical challenges, the main reason is that it requires physical insights about the emergent behavior, such as the notion of an appropriate order parameter.
An alternative way to identify phase transitions 
is through analyzing their quantum entanglement patterns, such as ``entanglement susceptibilities'' \cite{WK_entanglement_susceptibility}. 

For a system in a pure state, entanglement gives rise to effective mixed states of the subsystems, which have non-vanishing von Neumann entropy 
\cite{Calabrese2009, Eisert2010}. Given a system in the ground state $\ket{\Psi}$, the entanglement entropy between a subsystem, $A$, and the rest of the system, $B$ is the von Neumann entropy of the reduced density matrix,
\begin{equation}\label{eq:entanglement-entropy}
    \Scal_A=-\tr_A (\rho_A\log_2\rho_A),
\end{equation}
where $\rho_A=\tr_B\ket{\Psi}\bra{\Psi}$ is the reduced density matrix of the subsystem \cite{Bennett1996}. In general, for a gapped system, the entanglement entropy in the ground state is expected to follow 
an ``area law'', proportional to the boundary area between $A$ and $B$, with a non-universal prefactor \cite{Eisert2010}. The entanglement entropy is in general not extremal at critical points, therefore it cannot be easily utilized to locate phase transitions. However, additional universal corrections to the area law may arise at critical points that are extremal, with direct applicability to locate critical points. 
%
%
In particular, the ground states of a large class of gapless systems show a logarithmically divergent contribution to the entanglement entropy with a universal prefactor \cite{calabrese2004, Refael2009}. This ``corner contribution'' is due to the presence of geometric singularities in the boundary of the subregion, such as endpoints of an interval in $1d$ or sharp corners of a (hyper)cubic subsystem in higher dimensions \cite{Fradkin2006, Zaletel2011}. 
Specifically, for a $d-$dimensional system of linear size $L$ and subsystem of linear size $\ell\sim L$, the corner term is given by
\begin{equation}\label{eq:scr}
    \Scal_{cr}^{(d)}(\ell) \simeq b^{(d)}\ln \ell + \text{const,} 
\end{equation}
where $ b^{(d)}$ is universal, i.e., independent from the microscopic details, like the lattice type or disorder type, as discussed next.

\begin{figure}[ht]
    \centering
    \includegraphics[width=0.7\linewidth]{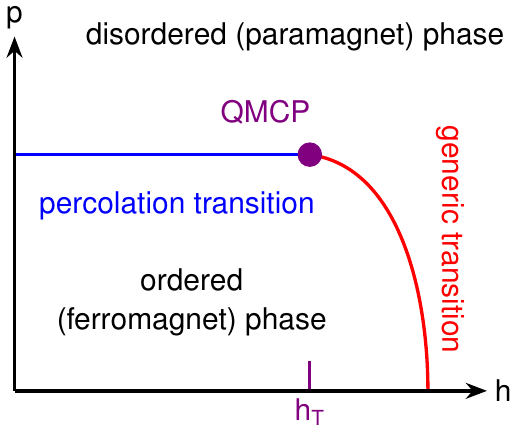}
    \vskip-0.2cm
    \caption{\textbf{Phase diagram of the RTFIM in $d\geq 2$}. The three studied quantum phase transitions indicated in different colors. QMCP stands for the quantum multi-critical point at the junction of the percolation and generic quantum critical transitions.}
    \label{fig:phase-diag}
\end{figure}

As a special case, in $1d$, where the area law is a constant, the von Neumann entropy of critical ground states is well known to diverge logarithmically with a universal prefactor. For conformally invariant clean systems, such as XY, XXZ, and Ising spin-$1/2$ chains, this prefactor depends on the central charge of the underlying conformal field theory \cite{vidal2003, calabrese2004, holzhey1994}. In the case of disordered $1d$ systems, such as the random antiferromagnetic Heisenberg model, the random XX chain, and the random transverse-field Ising model (RTFIM), the scaling form remains the same, with the central charge being replaced by an ``effective'' central charge \cite{Refael2009, laflorencie2005, Fisher1999}.

In higher dimensions, conformally invariant systems, such as the quantum dimer model \cite{Rokhsar1988}, free fermions \cite{Helmes2016}, and the quantum critical Ising model \cite{calabrese2004, Tagliacozzo2009}, again display a logarithmic prefactor of the corner term proportional to the central charge \cite{faulkner2016}. For non-conformally invariant systems with $d>1$, a subleading logarithmic term with a universal prefactor is likewise observed in the entanglement entropy, for both clean \cite{Singh2012, Humeniuk2012, Tagliacozzo2009, Song2011, Kallin2011} and disordered \cite{Kovacs2012-rtfim} models.
Therefore, in higher dimensional systems, instead of the full entanglement entropy, or the leading order area law contribution, the critical point and the corresponding universality class can be efficiently probed using the corner term. 

The focus of this study is the paradigmatic RTFIM on a diluted lattice, which, in two and higher dimensions, covers three distinct universality classes (figure~\ref{fig:phase-diag}). For sufficiently low transverse field strength, the critical behavior of the model at the bond dilution $p=p_c$ is governed by the classical percolation universality class. Below $p_c$, the generic quantum transition occurs by tuning the strength of quantum fluctuations via the transverse field parameter $h$. At the junction of these two transition lines of quantum critical points (QCPs) the quantum multi-critical point (QMCP) is found. 
Experimentally, the RTFIM can be realized in a number of compounds. One prominent example is the dipolar magnetic insulator ${\rm LiHo}_x{\rm Y}_{1-x}{\rm F}_4$ compound, a diluted relative of the compound ${\rm LiHoF}_4$ with a fraction of magnetic Ho atoms is replaced by nonmagnetic Y atoms \cite{Reich1990, Wu1991, Wu1993, Bitko1996, Brooke1999, Silevitch2010}. QMCPs also have particular experimental relevance, given that ferromagnetic QCPs are typically difficult to access experimentally, either because they change to first-order transitions or become buried within an intervening phase. QMCPs are expected to be more easily accessible, for instance see the recent experimental study of the compound Nb$_{1-y}$Fe$_{2+y}$ \cite{Friedemann2018}.



\begin{figure}[ht]
    \centering
    \includegraphics[width=\linewidth]{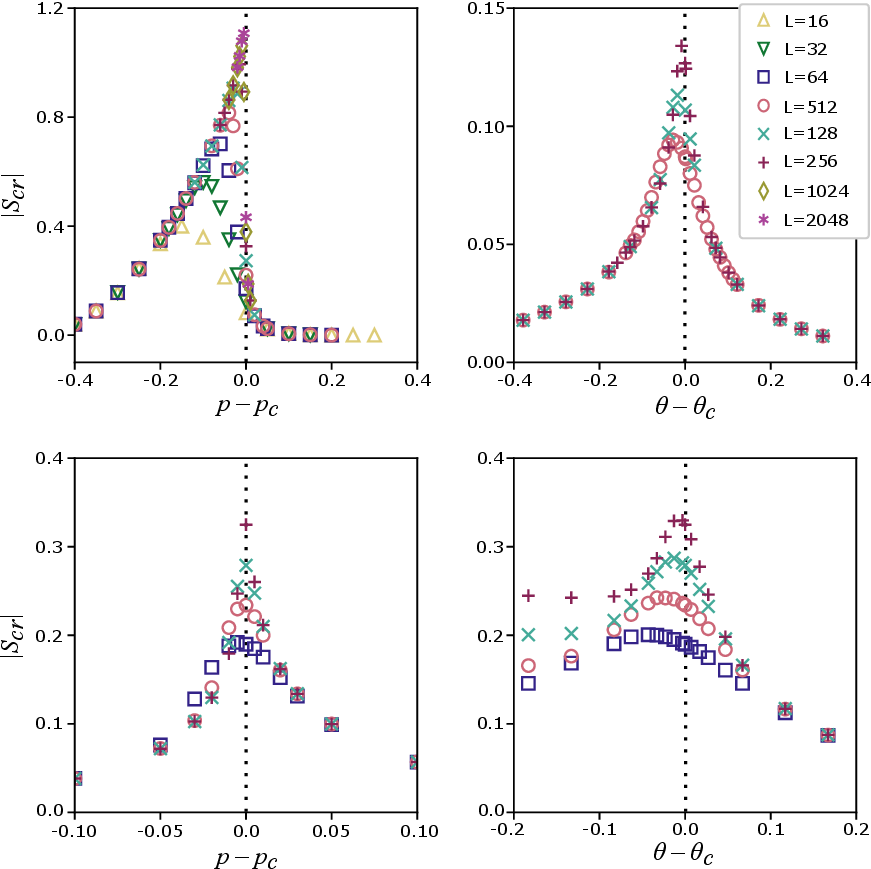}
     \vskip -.3cm
    \caption{\textbf{Locating the phase transition via the corner contribution.} The corner contribution to the entanglement entropy for a regular square shows a sharp peak at a quantum phase transition, as a function of the control parameter ($p$ or $\theta$). Clockwise from top left: 
    percolation QCP\cite{Kovacs2012-percolation}, generic QCP \cite{Kovacs2012-rtfim}, QMCP ($h$-direction), QMCP ($p$-direction) \cite{Kovacs2024-qmcp}.}
    \label{fig:grp}
\end{figure}

It has been recently shown that the corner term to the entanglement entropy exists, has a universal logarithmic form, and can serve as an entanglement susceptibility for all three of these universality classes \cite{Yu2008, Kovacs2012-rtfim, Kovacs2012-percolation, Kovacs2024-qmcp}, at least for square subsystems, as shown in figure~\ref{fig:grp}. 
Although being universal, the corner term must depend on the shape (i.e., opening angle) of the subsystem corners \cite{WK-corner-entanglement}, however this has not been explored for the generic QCP and the QMCP. 

\begin{figure}[ht]
    \centering
    \includegraphics[width=\linewidth]{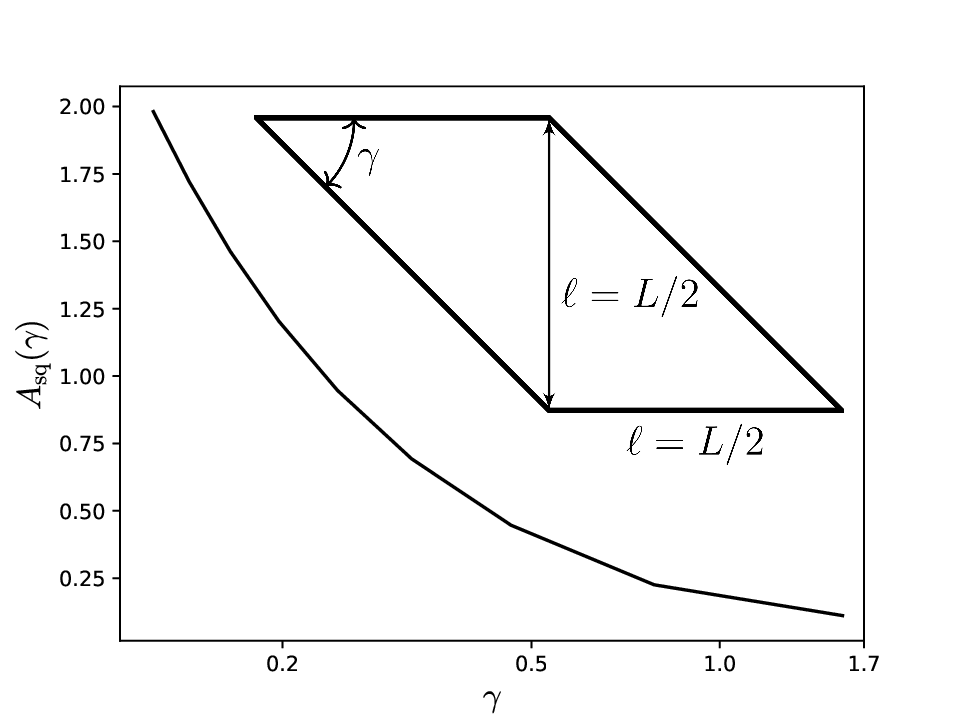}
    \vskip -0.3cm
    \caption{\textbf{Angle dependence at the percolation QCP}. Conformally invariant systems follow the Cardy-Peschel formula for the angle dependence in equation~(\ref{eq:shape-factor}) \cite{cardy1988}, as shown here for percolation, equation~(\ref{eq:cp}). The inset shows an example of the studied sheared square subsystem geometry. The volume of the subsystem is $\ell^2=L^2/4$, equal to the volume of a square subsystem with the same linear extent, and its interior angles are $\gamma$ and $\pi-\gamma$.}
    \label{fig:shear-square}
\end{figure}

For conformally invariant $2d$ systems, the shape-dependence of the corner term is expected to follow the celebrated Cardy-Peschel formula \cite{cardy1988}, including 
the percolation QCP \cite{Kovacs2012-percolation, Kovacs2014-pottscorners}. 
For example, a sheared square parallelogram 
of a (smaller) interior angle $\gamma$ (figure \ref{fig:shear-square}) yields
\begin{equation}\label{eq:cp}
    b(\gamma) = c'(1) A_{\text{sq}}(\gamma),
\end{equation}
where $c'(1)=\frac{5\sqrt{3}}{4\pi}$ is the derivative of the central charge $c(Q)$ of the $Q$-state Potts model, at $Q\rightarrow1$ for percolation, and
\begin{equation}\label{eq:shape-factor}
    A_{\text{sq}}(\gamma) = \frac{1}{12}\left(4-  \frac{\pi}{\gamma} - \frac{\pi}{\pi-\gamma} - \frac{\pi}{\pi + \gamma} - \frac{\pi}{2\pi - \gamma} \right)
\end{equation}
is the shape-dependent factor.
As in this case the shape-dependence is completely dictated by conformal invariance, different subsystem geometries do not offer new information as compared to a regular square. The goal of our investigations is to determine whether, for the generic QCP and QMCP, the shape-dependence of the corner contribution obeys a similar form to the conformal prediction.
Earlier studies of the RTFIM in one-dimensional chains have found that conformal predictions are generally a close fit, for instance, for the magnetization profile; however, the discrepancies with numerical data are nontrivial \cite{Rieger1997, RTFIC2008}.

In this paper, we present a comprehensive picture of the universal features and shape-dependence of quantum entanglement for the full range of QCPs of the RTFIM in 2 dimensions. Our results indicate that --- unlike in conformally invariant systems --- considering different subsystem shapes is informative on the emergent correlation structure for the generic QCP and QMCP, beyond what is probed by a square subsystem.

\section*{Model and Method}

\begin{figure*}[ht]
    \includegraphics[width=\linewidth]{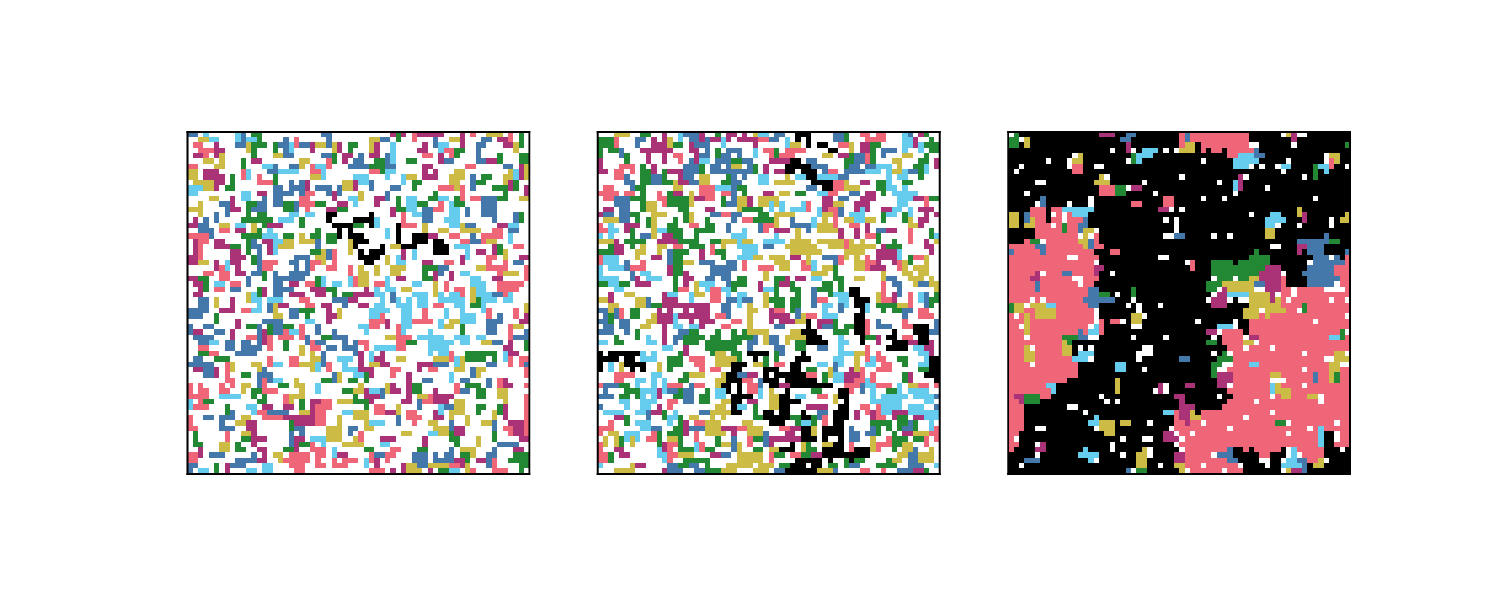}
    \vskip -1.7cm
    \caption{\textbf{Critical and multi-critical ground state configurations.} The ground state of the RTFIM consists of independent GHZ spin clusters in all three universality classes, as shown here for $64\times64$ systems. (Left: generic QCP, fixed-$h$ disorder; middle: QMCP, fixed-$h$ disorder; right: percolation QCP, bond percolation.) The largest cluster in each sample is shown in black. The generic and multi-critical clusters are generally geometrically disconnected, unlike the percolation clusters.}
    \label{fig:rtfim-clusters}
\end{figure*}

The RTFIM is defined by the Hamiltonian
\begin{equation}\label{eq:rtfim-hamiltonian}
    \Ham = -\sum_{\langle ij\rangle} J_{ij}\sigma_i^x\sigma_j^x -\sum_i h_i \sigma_i^z,
\end{equation}
where the Pauli matrices $\sigma_{i}^{x,z}$ represent spins on the sites $i$ of a $d-$dimensional square lattice with periodic boundary conditions. The lattice is diluted with some bond dilution probability $p$, such that each link between sites in the lattice is independently removed with probability $p$. The spins interact through the nearest-neighbor couplings $J_{ij}$, and each site is also subject to a transverse field $h_i$, each of which are independent random numbers taken from the distributions $p(J)$ and $q(h)$, respectively. Following refs. \cite{Kovacs2009-rtfim, Kovacs2010-2d-sdrg, Kovacs2011-3d-sdrg, Kovacs2011-sdrg, Kovacs2022-qmcp, Kovacs2024-qmcp}, in order to test the universality of the results, we have used two disorder distributions; for box-$h$ disorder the fields are uniformly distributed in $[0,h_b]$, for fixed-$h$ disorder $h_i=h_f$ at every site, and in both cases $p(J)$ is uniformly distributed in $[0,1]$. The quantum control parameter is defined as $\theta=\ln h_b$ or $\theta=\ln h_f$ for box-$h$ or fixed-$h$ disorder, respectively. In $2d$, the percolation QCP is found at the bond percolation threshold $p_c=0.5$, with the QMCP located at the terminus of this line at $\theta_T=0.783$ $(-0.481)$ for box-$h$ (fixed-$h$) disorder \cite{Kovacs2022-qmcp}. The generic QCP is located, for $p=0$, at $\theta_c=1.6784$ $(-0.17034)$ for box-$h$ (fixed-$h$) disorder \cite{Kovacs2012-rtfim}.



The presence of quenched disorder in the RTFIM facilitates a numerical treatment using the strong-disorder renormalization group (SDRG) method \cite{Ma1976, Dasgupta1980}, as the disorder fluctuations dominate over the quantum fluctuations. This is due to the fact that the critical properties of the model are controlled by an infinite-disorder fixed point, where the strength of the disorder diverges in renormalization, and thus the SDRG results are asymptotically exact in the vicinity of the critical point \cite{Fisher1999, Pandey2023}.
The SDRG is an iterative perturbative calculation, in which the largest (local) terms in the Hamiltonian in equation~(\ref{eq:rtfim-hamiltonian}) are successively eliminated and new effective Hamiltonians are generated, according to the following perturbative decimation rules. 

When a coupling $J_{ij}$ is decimated, the two sites $i,j$ are aggregated to form an effective spin cluster with an effective transverse field $h'\approx h_ih_j/J_{ij}$. New couplings are then formed between the cluster and all neighbors $a$ of $i,j$, the renormalized value of which is given by the so-called ``maximum rule'' $J_a'=\max(J_{ai}, J_{aj})$. On the other hand, when a transverse field $h_i$ is decimated, the site is removed and new effective couplings are formed between all of site $i$'s neighbors $a, b$, given by $J'_{ab}\approx J_{ai}J_{bi}/h_i$. If sites $a$ and $b$ are already connected by a nonzero coupling $J_{ab}$, then the maximum rule is similarly applied, taking $\max(J_{ab}', J_{ab})$. The maximum rule is expected to be a good approximation close to an IDFP, where the distribution of the logarithmic couplings is extremely broad, meaning that the larger coupling can be orders of magnitudes larger than the other one.
In practice, the ground state of the RTFIM is calculated using an efficient numerical implementation of the SDRG method \cite{Kovacs2009-rtfim, Kovacs2010-2d-sdrg, Kovacs2011-3d-sdrg, Kovacs2011-sdrg}.

At each iteration of the procedure, the number of effective sites is reduced by one. Once all degrees of freedom have been decimated, the resulting ground state of the system is a collection of independent ferromagnetic clusters (figure \ref{fig:rtfim-clusters}), each in a GHZ state $\frac{1}{\sqrt{2}}(\ket{\uparrow \uparrow \dots \uparrow}+\ket{\downarrow \downarrow \dots \downarrow})$. Each such cluster contributes $\log_2 2=1$ to the entanglement entropy of a subsystem if the cluster has at least one site inside and one site outside of the subsystem, otherwise it plays no role \cite{Refael2009}. Thus, calculating entanglement entropy for the RTFIM is equivalent to a cluster counting problem. Notably, the clusters are in general geometrically disconnected at the generic QCP, unlike in the case of percolation \cite{Kovacs2012-percolation, Helen-percolation}, as illustrated in figure \ref{fig:rtfim-clusters}.

\section*{Results}

\begin{figure}
    \centering
    \includegraphics[width=\linewidth]{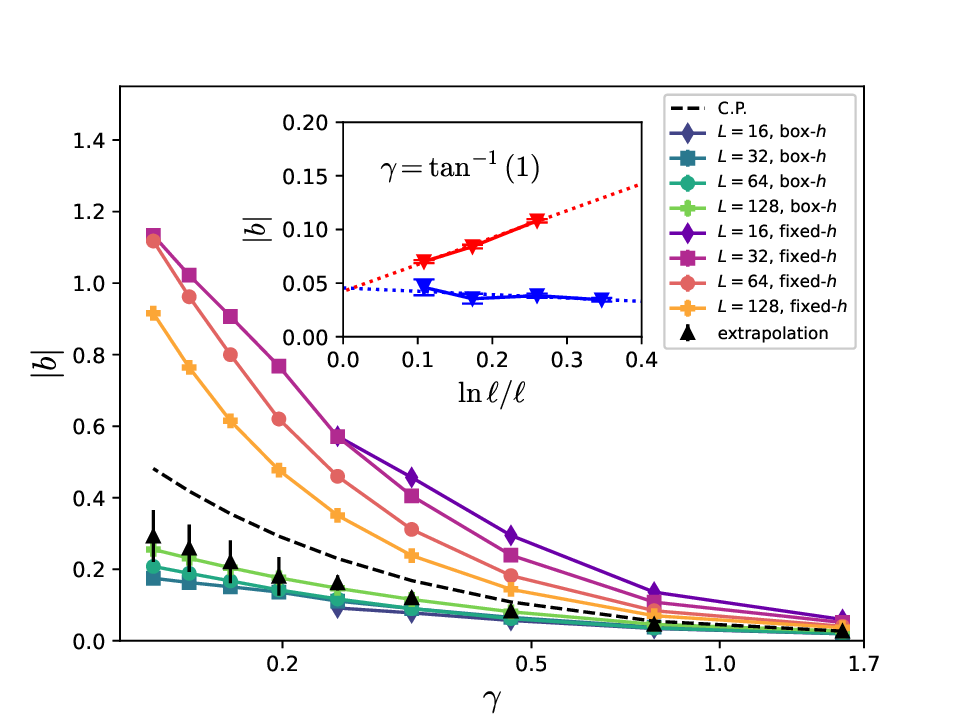}
    \includegraphics[width=\linewidth]{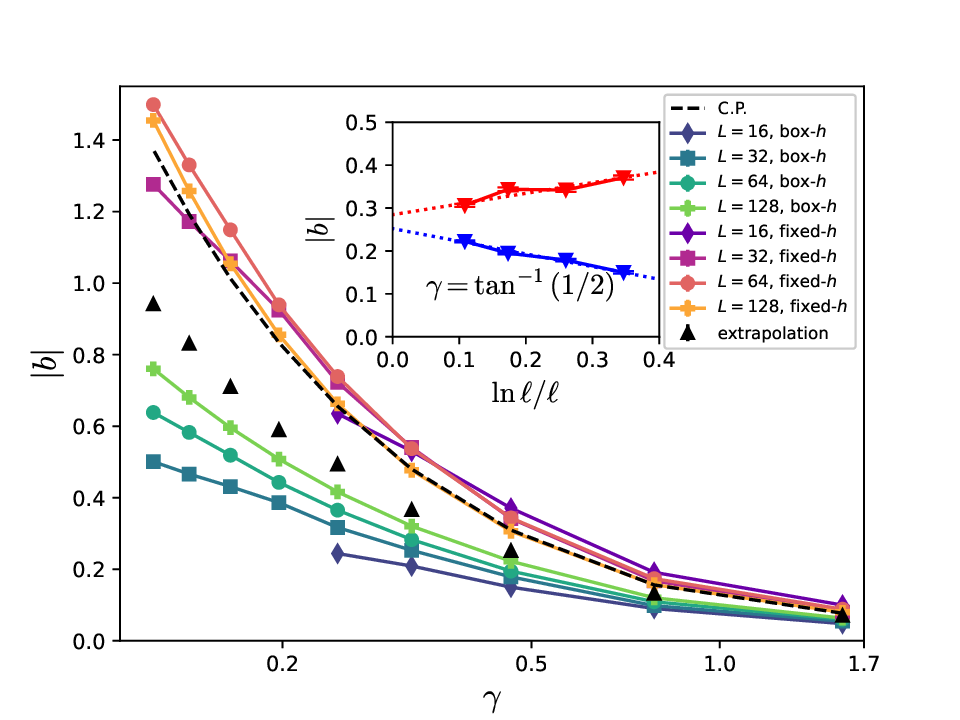}
    \vskip -0.25cm
    \caption{\textbf{Shape-dependence at the generic QCP and QMCP.} Estimates for the generic QCP (top) and QMCP (bottom) prefactor $-b$ for sheared squares as a function of $\ln\gamma$. At each angle, the numerical results converge to the same asymptotic values as indicated by triangles and the corresponding standard error.
    The insets show the infinite size extrapolation at $\gamma=\tan^{-1}(1)$ (top) and $\gamma=\tan^{-1}(1/2)$ (bottom).
    For reference, the dashed line (C.P.) shows the conformal prediction based on the Cardy-Peschel formula, with an effective $c'(1)$ calculated using the regular square data. 
    }
    \label{fig:qcp-shape-dep}
\end{figure}

Using the geometric method (see \href{appendixA}{Appendix A.}, \cite{Kovacs2012-percolation, Kovacs2012-rtfim, Kovacs2024-qmcp}), we calculated the average corner contribution in equation~(\ref{eq:scr}) for sheared square subsystems with shear angle $\gamma=\tan^{-1}(1/n)$, $n=0,1,2,\dots,8$, in systems of size $L=8, 16, 32, 64, 128$, for $10,000$ realizations at each size and disorder type. We used periodic boundary conditions (PBC) in all directions. We then obtained size-dependent estimates for the prefactor for each subsystem geometry using two-point fits by comparing the average corner contribution at size $L$ and $2L$, and performed an extrapolation in $\ln \ell/\ell$ at each angle $\gamma$. The results at the generic QCP and QMCP are presented in figure \ref{fig:qcp-shape-dep}. 

\begin{figure}
    \centering
    \includegraphics[width=\linewidth]{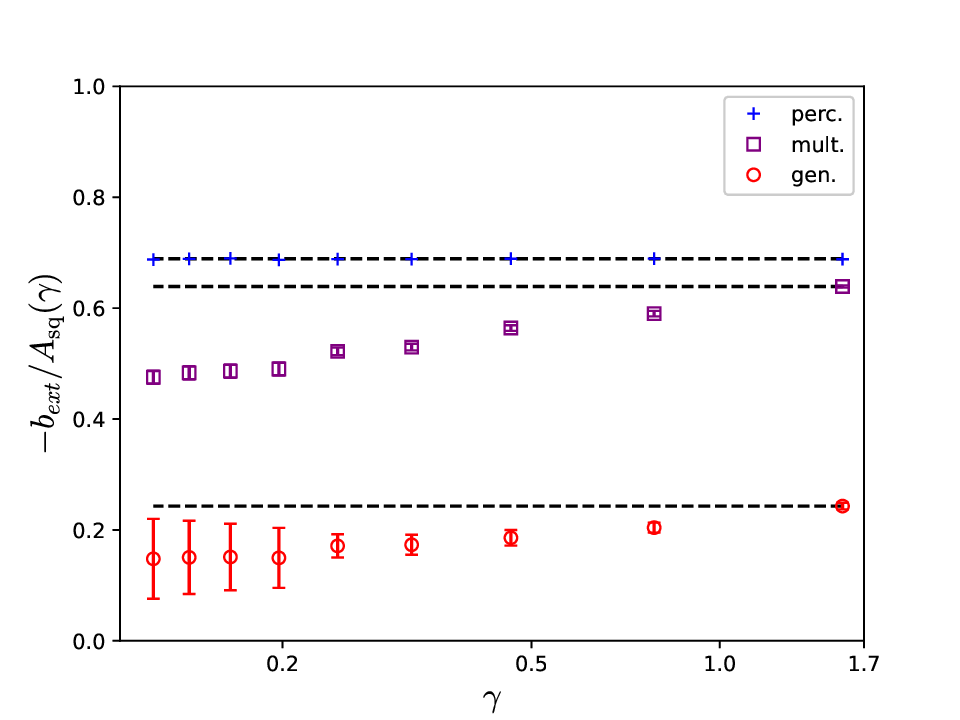}
        \vskip -0.25cm
    \caption{\textbf{Comparison to the conformal predictions for sheared squares.} The ratio of extrapolated values of $-b$ and $A_{\text{sq}}(\gamma)$. The error bars represent the standard error, which at most points is smaller than the size of the markers. The horizontal lines indicate the effective $c'(1)$ for each case, calculated using the regular square data.}
    \label{fig:b-ext}
\end{figure}

We found strong finite-size effects for both disorder types, however, displaying an opposite sign, increasing for box-$h$ disorder, while decreasing for fixed-$h$ disorder, in line with what was observed for square subsystems in the literature \cite{Kovacs2012-rtfim, Kovacs2024-qmcp}.
Yet, the extrapolated prefactors for each disorder type were found to be generally in agreement, supporting the universality (i.e. disorder-independence) of the results. 
Dividing our results at the generic QCP and the QMCP by the geometric factor $A_{\rm sq}(\gamma)$ in equation~(\ref{eq:shape-factor}), the shape dependence of $b$ at these points does not appear to obey the conformal prediction as shown in figure \ref{fig:b-ext}. This indicates that the generic quantum critical and quantum multi-critical points, unlike critical percolation, are not conformally invariant. As the shape-dependence of $b$ is nontrivial at these points, different subsystem shapes capture different information about the underlying entanglement patterns. Additionally, the prefactors $b(\gamma)$ at the QMCP lie between the values at the percolation and generic QCPs, which is consistent with the overall trend that the multi-critical properties interpolate between the generic and percolation critical properties \cite{Kovacs2022-qmcp, Kovacs2024-qmcp}.

\begin{figure}[ht]
    \centering
    \includegraphics[width=0.48\linewidth]{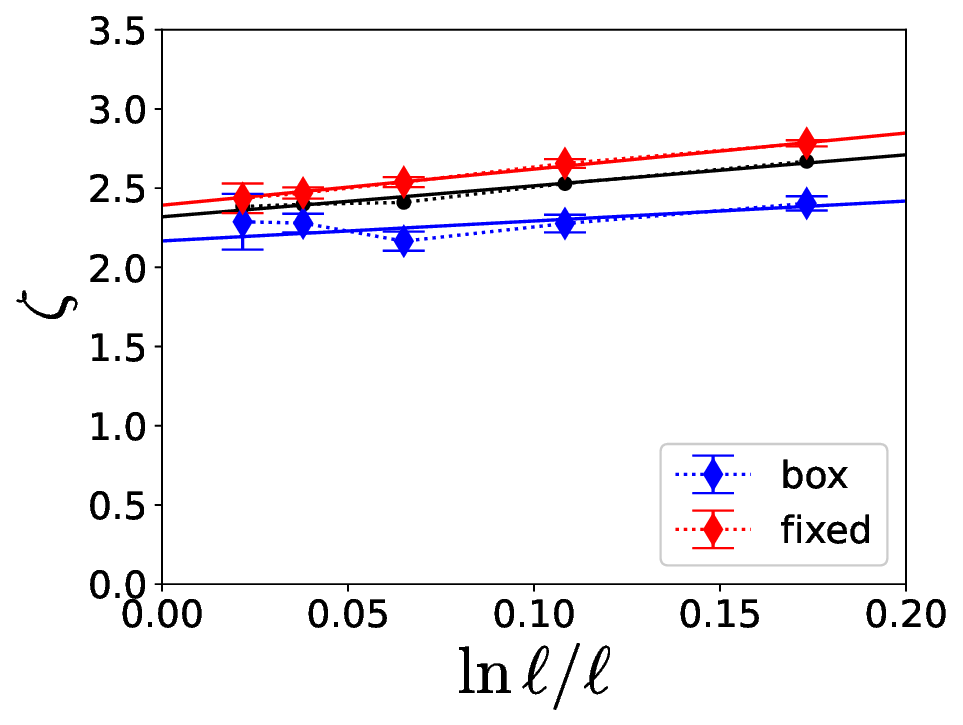}
    \includegraphics[width=0.48\linewidth]{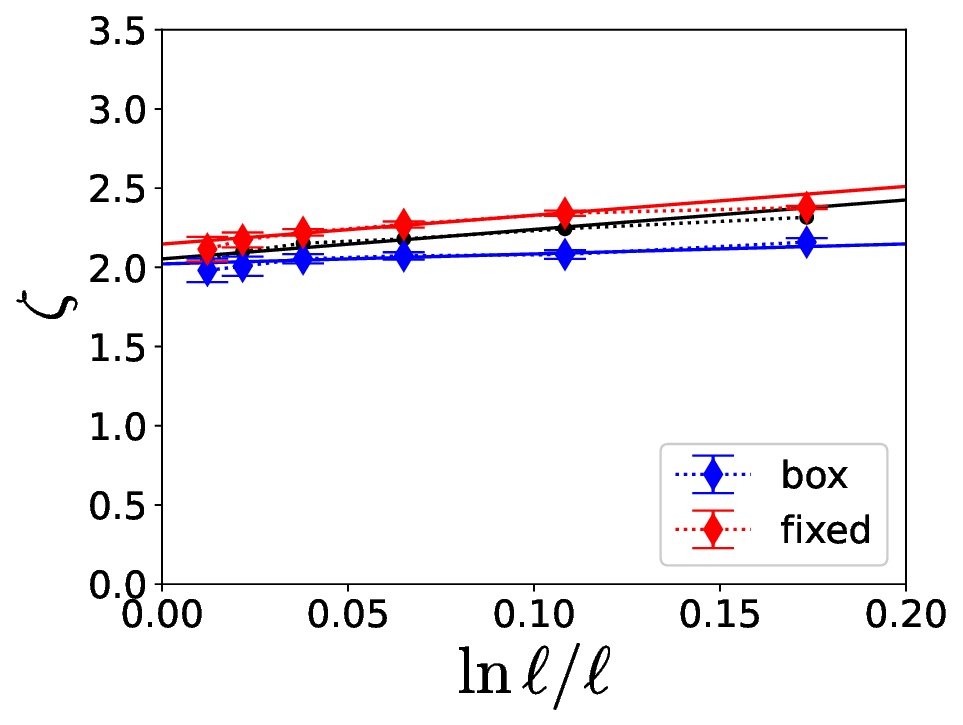}

    \includegraphics[width=0.48\linewidth]{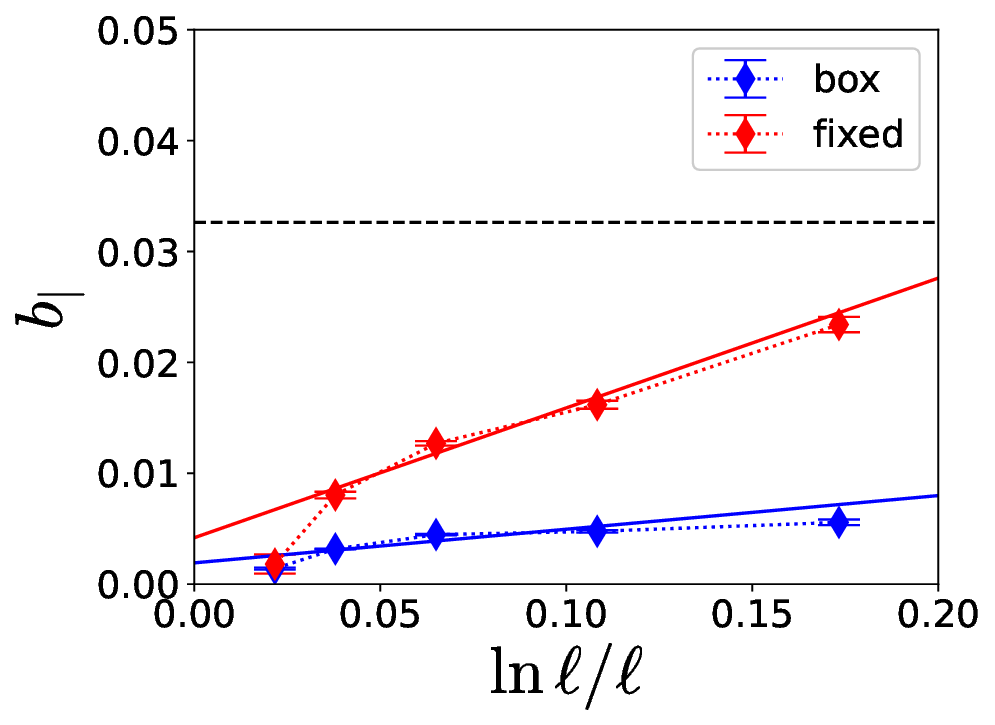}
    \includegraphics[width=0.48\linewidth]{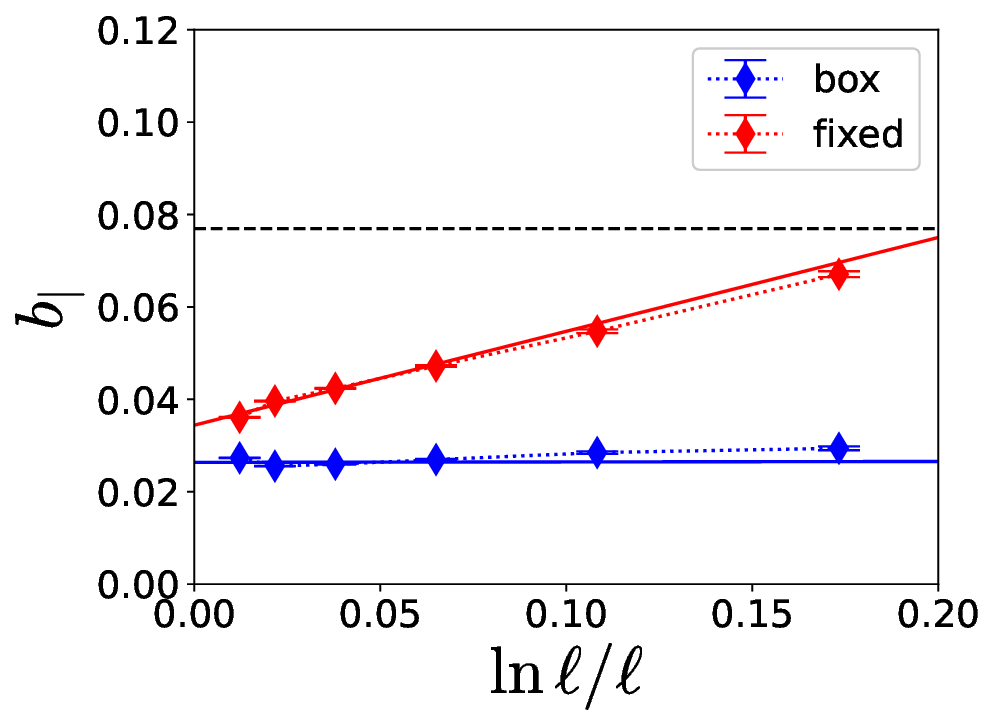}
    \vskip -0.3cm
    \caption{\textbf{Results for skeletal entanglement using line segments.} Finite-size scaling of the gap size exponent $\zeta$ (top) and the corner contribution prefactor $b$ for line segments, see equation~(\ref{eq:double-sum}) (bottom). Left: generic QCP, $\zeta=2.3(4)$, $b=0.003(1)$. Right: QMCP, $\zeta=2.1(4)$, $b=0.30(4)$. The error bars indicate the standard error, which at most points is smaller than the size of the markers. The solid black lines represent the average (weighted with uncertainty) of the two disorder distributions, the intercept of which gives $\zeta$. The dashed black lines indicate the expected value of $b$ based on square subystems if the angle dependence follows the conformally invariant results given by the Cardy-Peschel formula.}
    \label{fig:line-zeta}
\end{figure}

In addition to sheared squares, we have also investigated the ``skeletal'' entanglement properties of zero-volume line-segment subsystems in the RTFIM (again with PBC). We used the same set of samples generated for the sheared square calculation and an additional 10,000 realizations of each disorder type at $L=256$. In conformally invariant systems, the entanglement entropy of a line segment of length $\ell$ typically scales with the length of the line according to an area law, but for finite $\ell < L$ there is a logarithmic corner contribution with a universal prefactor \cite{Helen-percolation, WK-skeletal-entanglement, Yu2008}. The prefactor has been found to depend only on whether and at what angle the endpoints of the line segment intersect with the (free) surface of the system. 
In the simplest geometry of having a line segment fully in the bulk, the corner contribution is related to a quantity $n(s)$ known as either the ``gap-size" or ``chord-length" statistics. For a given sample, $n(s)$ is the number of times that a distance gap of size $s$, where $1\leq s\leq L/2$ due to PBC, occurs between sites of the same cluster. Using this, the average corner contribution within a sample of (horizontal) line segments of length $\ell=L/2$ is \cite{Kovacs2012-rtfim,Helen-percolation}
\begin{equation}\label{eq:double-sum}
    \Scal^{(1)}_{cr}(\ell=L/2) = \frac{1}{L^2}\sum_{i=1}^\ell\sum_{s=i}^{L/2}n(s).
\end{equation}
Both in the case of percolation, and in the $1d$ RTFIM, the gap-size statistics are known to follow $n(s) \propto s^{-\zeta}$, with $\zeta=2$, in line with the logarithmic form of the corner contribution. 

Following the methodology described in \cite{Helen-brain}, we have used finite-size scaling to extrapolate the value of the exponent $\zeta$ as $L\rightarrow\infty$, at both the generic QCP and QMCP. Performing an extrapolation in $\ln\ell/\ell$ we find $\zeta=2.3(4)$ at the generic QCP and $\zeta=2.1(4)$ at the QMCP (figure \ref{fig:line-zeta}). $\zeta>2$ would indicate a corner contribution that asymptotically saturates to a size-independent constant for large $L$, yielding $b=0$. 
Therefore, we have applied the formula~(\ref{eq:double-sum}) to calculate the line segment corner contribution and extrapolate the prefactor $b$ (figure \ref{fig:line-zeta}). These extrapolated prefactors differ significantly from the values predicted by conformal invariance in both cases. This further indicates that neither the generic quantum critical nor quantum multi-critical points are conformally invariant. 
While the QMCP result is significantly different from zero, clearly indicating $\zeta=2$, the results for the generic QCP are at least an order of magnitude smaller, consistent with $b=0$ and $\zeta>2$. Note that as a consequence of the geometric method and equation~(\ref{eq:double-sum}), the corner term of line segments must be non-negative for each sample and each subsystem location. This implies that even if the value of $b$ is asymptotically 0, the two disorder realizations would not be able to approach it from opposite directions, leading to a systematic bias in the estimates compared to the other cases studied here.

Altogether, we conjecture that the QMCP is again interpolating between the behavior of the percolation and generic QCPs in the following way:
while at the percolation QCP conformal invariance holds with $\zeta=2$, at the generic QCP conformal invariance does not apply with $\zeta>2$ ($\zeta\approx 2.3$) and $b=0$, with the QMCP breaking conformal invariance, but with $\zeta=2$ and $b=0.030(4)$. As future work, it would be of interest to explore the nine other line segment configurations that have been studied for $2d$ percolation, as some of these led to an order of magnitude larger values of $b$ \cite{Helen-percolation}. This can lead to improved statistics and a better distinction between 0 and non-zero results. 

\section*{Discussion}

We have studied the shape-dependence of entanglement entropy at the full range of quantum critical points of a paradigmatic disordered interacting quantum system, the RTFIM in 2 dimensions. The emerging picture is expected to be fairly generic: although the entanglement entropy of a subsystem scales according to an area law, at a critical point there is a logarithmic correction with a universal prefactor $b$. While at the percolation QCP, the shape-dependence of $b$ follows entirely from conformal invariance, at the generic QCP and QMCP, and likely at other generic disordered fixed points, the shape-dependence of $b$ is not given by the conformal prediction and is therefore informative.

One future direction is to extend our results to the RTFIM in 3 dimensions, where the logarithmic prefactor has been found to be universal for a regular cubic subsystem \cite{Kovacs20143dperc, Kovacs2012-rtfim, Kovacs2024-qmcp}, but the shape-dependence has not been investigated systematically across all critical regimes. Based on our results, we anticipate that new shapes will in general provide new information on the underlying universality class. Note that in 3 dimensions $2d$ subsystems provide an alternative, intermediate form of skeletal entanglement besides $1d$ line segments. In addition, numerical studies of skeletal entanglement for percolation clusters in $3d$ have revealed that line segments contained entirely on a (free) surface have a non-singular logarithmic component, along with a gap size exponent $\zeta>2$ on the cube faces \cite{Helen-percolation}. As we already observe $\zeta>2$ and vanishing $b$ for line segments in the bulk at the generic QCP, we expect similar results in higher dimensions, including on surfaces. Given the trend of behavior at the QMCP to interpolate between the percolation and generic QCPs, it is also expected that at the QMCP line segments on the surface would not give a singular corner contribution.

Also of interest is the question of the dynamical behavior of the entanglement entropy, after a sudden change in the Hamiltonian parameters. In $1d$, when the global quench is performed at the critical point, the entanglement entropy is found to grow ultra-slowly, as $\Scal(t)\sim \ln\ln t$ \cite{Igloi2012-dynamics}. In higher dimensions, the corner contribution is expected to obey a similar time dependence.

In addition, studying the multipartite entanglement between two non-adjacent subsystems appears to be a promising avenue, capturing additional universal aspects of entanglement patterns \cite{Parez2024, Wang2025, Lyu2025}. 
Using entanglement negativity and mutual information to quantify multipartite entanglement, in the appropriate scaling limit both of these measures have been found to be universal in the $1d$ RTFIM \cite{Zou2022-multipartite}. 
In higher dimensions, the entanglement entropy of a single subsystem is in general not universal, with only the corner contribution featuring a universal prefactor. However,  multipartite measures of entanglement can be entirely universal, serving again as a notion of entanglement susceptibility. Unlike in one-dimensional systems, in higher dimensions, in addition to the distance being a tunable parameter, the shape of each subsystem \cite{Eisler2016} and their relative orientation can also impact the results in universal ways.

Finally, there are also classical systems, which are found in the same universality classes as quantum critical points and to which these results can be extended. Notably, the disordered contact process of infection spreading on a lattice falls into the three universality classes of the RTFIM  studied here in both 2$d$ and 3$d$ \cite{Vojta2009, Vojta2012, Hooyberghs2003, Hooyberghs2004, Kovacs2022-qmcp, luzzatto2025}, and exhibits a similar structure of spatially disconnected correlation clusters \cite{kovacs2020-contactprocess}. 
%
We expect that the cluster counting methodology described here, including the geometric trick, can be adapted to obtain accurate corner contributions based on either all the correlation clusters or even if only the 
largest cluster is considered. 
Such out-of-equilibrium classical measurements would serves as independent and deep tests of the entanglement results presented here. 

\section*{Acknowledgements}
This work was supported by the National Science Foundation under Grant No.~PHY-2310706 of the QIS program in the Division of Physics, supplemented by the MPS 
AGEP-GRS. This research was supported in part through the computational resources and staff contributions provided for the Quest high performance computing facility at Northwestern University, which is jointly supported by the Office of the Provost, the Office for Research, and Northwestern University Information Technology.

\section*{Code and Data Availability}
The cluster counting codes used in this project can be found in the repository \href{https://github.com/natalielove2029-lab/shape-dependence}{https://github.com/natalielove2029-lab/shape-dependence}. The raw data files can be found in the repository \href{https://github.com/natalielove2029-lab/shape-dep-data}{https://github.com/natalielove2029-lab/shape-dep-data}.

\bibliography{shapedep}

@article{RTFIC2008,
  title = {Density of critical clusters in strips of strongly disordered systems},
  author = {Karsai, M. and Kov\'acs, I. A. and Angl\`es d'Auriac, J-Ch. and Igl\'oi, F.},
  journal = {Phys. Rev. E},
  volume = {78},
  issue = {6},
  pages = {061109},
  numpages = {9},
  year = {2008},
  month = {Dec},
  publisher = {American Physical Society},
  doi = {10.1103/PhysRevE.78.061109},
  url = {https://link.aps.org/doi/10.1103/PhysRevE.78.061109}
}

@article{Rieger1997,
  title = {Density Profiles in Random Quantum Spin Chains},
  author = {Igl\'oi, Ferenc and Rieger, Heiko},
  journal = {Phys. Rev. Lett.},
  volume = {78},
  issue = {12},
  pages = {2473--2476},
  numpages = {0},
  year = {1997},
  month = {Mar},
  publisher = {American Physical Society},
  doi = {10.1103/PhysRevLett.78.2473},
  url = {https://link.aps.org/doi/10.1103/PhysRevLett.78.2473}
}

@misc{luzzatto2025,
      title={Multicritical Infection Spreading}, 
      author={Leone V. Luzzatto and Juan Felipe Barrera López and István A. Kovács},
      year={2025},
      eprint={2508.20895},
      archivePrefix={arXiv},
      primaryClass={cond-mat.stat-mech},
      url={https://arxiv.org/abs/2508.20895}, 
}

@article{Wang2025,
	title = {Entanglement microscopy and tomography in many-body systems},
	volume = {16},
	issn = {2041-1723},
	url = {https://www.nature.com/articles/s41467-024-55354-z},
	doi = {10.1038/s41467-024-55354-z},
	number = {1},
	urldate = {2025-07-03},
	journal = {Nat Commun},
	author = {Wang, Ting-Tung and Song, Menghan and Lyu, Liuke and Witczak-Krempa, William and Meng, Zi Yang},
	year = {2025},
	pages = {96},
}

@article{Parez2024,
  title = {Entanglement negativity between separated regions in quantum critical systems},
  author = {Parez, Gilles and Witczak-Krempa, William},
  journal = {Phys. Rev. Res.},
  volume = {6},
  issue = {2},
  pages = {023125},
  numpages = {6},
  year = {2024},
  month = {May},
  publisher = {American Physical Society},
  doi = {10.1103/PhysRevResearch.6.023125},
  url = {https://link.aps.org/doi/10.1103/PhysRevResearch.6.023125}
}

@article{Lyu2025,
  title = {Multiparty entanglement microscopy of quantum {I}sing models in one, two, and three dimensions},
  author = {Lyu, Liuke and Song, Menghan and Wang, Ting-Tung and Meng, Zi Yang and Witczak-Krempa, William},
  journal = {Phys. Rev. B},
  volume = {111},
  issue = {24},
  pages = {245108},
  numpages = {16},
  year = {2025},
  month = {Jun},
  publisher = {American Physical Society},
  doi = {10.1103/PhysRevB.111.245108},
  url = {https://link.aps.org/doi/10.1103/PhysRevB.111.245108}
}

@article{Singh2012,
  title = {Thermodynamic singularities in the entanglement entropy at a two-dimensional quantum critical point},
  author = {Singh, Rajiv R. P. and Melko, Roger G. and Oitmaa, Jaan},
  journal = {Phys. Rev. B},
  volume = {86},
  issue = {7},
  pages = {075106},
  numpages = {5},
  year = {2012},
  month = {Aug},
  publisher = {American Physical Society},
  doi = {10.1103/PhysRevB.86.075106},
  url = {https://link.aps.org/doi/10.1103/PhysRevB.86.075106}
}

@article{Humeniuk2012,
  title = {Quantum {M}onte {C}arlo calculation of entanglement {R}\'enyi entropies for generic quantum systems},
  author = {Humeniuk, Stephan and Roscilde, Tommaso},
  journal = {Phys. Rev. B},
  volume = {86},
  issue = {23},
  pages = {235116},
  numpages = {8},
  year = {2012},
  month = {Dec},
  publisher = {American Physical Society},
  doi = {10.1103/PhysRevB.86.235116},
  url = {https://link.aps.org/doi/10.1103/PhysRevB.86.235116}
}

@article{Yu2008,
  title = {Entanglement entropy in the two-dimensional random transverse field {I}sing model},
  author = {Yu, Rong and Saleur, Hubert and Haas, Stephan},
  journal = {Phys. Rev. B},
  volume = {77},
  issue = {14},
  pages = {140402},
  numpages = {4},
  year = {2008},
  month = {Apr},
  publisher = {American Physical Society},
  doi = {10.1103/PhysRevB.77.140402},
  url = {https://link.aps.org/doi/10.1103/PhysRevB.77.140402}
}

@article{Reich1990,
  title = {Dipolar magnets and glasses: Neutron-scattering, dynamical, and calorimetric studies of randomly distributed {I}sing spins},
  author = {Reich, D. H. and Ellman, B. and Yang, J. and Rosenbaum, T. F. and Aeppli, G. and Belanger, D. P.},
  journal = {Phys. Rev. B},
  volume = {42},
  issue = {7},
  pages = {4631--4644},
  numpages = {0},
  year = {1990},
  month = {Sep},
  publisher = {American Physical Society},
  doi = {10.1103/PhysRevB.42.4631},
  url = {https://link.aps.org/doi/10.1103/PhysRevB.42.4631}
}

@article{Wu1991,
  title = {From classical to quantum glass},
  author = {Wu, Wenhao and Ellman, B. and Rosenbaum, T. F. and Aeppli, G. and Reich, D. H.},
  journal = {Phys. Rev. Lett.},
  volume = {67},
  issue = {15},
  pages = {2076--2079},
  numpages = {0},
  year = {1991},
  month = {Oct},
  publisher = {American Physical Society},
  doi = {10.1103/PhysRevLett.67.2076},
  url = {https://link.aps.org/doi/10.1103/PhysRevLett.67.2076}
}

@article{Wu1993,
  title = {Quenching of the nonlinear susceptibility at a {T=0} spin glass transition},
  author = {Wu, Wenhao and Bitko, D. and Rosenbaum, T. F. and Aeppli, G.},
  journal = {Phys. Rev. Lett.},
  volume = {71},
  issue = {12},
  pages = {1919--1922},
  numpages = {0},
  year = {1993},
  month = {Sep},
  publisher = {American Physical Society},
  doi = {10.1103/PhysRevLett.71.1919},
  url = {https://link.aps.org/doi/10.1103/PhysRevLett.71.1919}
}

@article{Brooke1999,
author = {J. Brooke  and D. Bitko  and T. F.  and G. Aeppli },
title = {Quantum Annealing of a Disordered Magnet},
journal = {Science},
volume = {284},
number = {5415},
pages = {779-781},
year = {1999},
doi = {10.1126/science.284.5415.779},
URL = {https://www.science.org/doi/abs/10.1126/science.284.5415.779}
}

@article{Hooyberghs2004,
  title = {Absorbing state phase transitions with quenched disorder},
  author = {Hooyberghs, Jef and Igl\'oi, Ferenc and Vanderzande, Carlo},
  journal = {Phys. Rev. E},
  volume = {69},
  issue = {6},
  pages = {066140},
  numpages = {16},
  year = {2004},
  month = {Jun},
  publisher = {American Physical Society},
  doi = {10.1103/PhysRevE.69.066140},
  url = {https://link.aps.org/doi/10.1103/PhysRevE.69.066140}
}

@article{Hooyberghs2003,
  title = {Strong Disorder Fixed Point in Absorbing-State Phase Transitions},
  author = {Hooyberghs, Jef and Igl\'oi, Ferenc and Vanderzande, Carlo},
  journal = {Phys. Rev. Lett.},
  volume = {90},
  issue = {10},
  pages = {100601},
  numpages = {4},
  year = {2003},
  month = {Mar},
  publisher = {American Physical Society},
  doi = {10.1103/PhysRevLett.90.100601},
  url = {https://link.aps.org/doi/10.1103/PhysRevLett.90.100601}
}

@article{Vojta2009,
  title = {Infinite-randomness critical point in the two-dimensional disordered contact process},
  author = {Vojta, Thomas and Farquhar, Adam and Mast, Jason},
  journal = {Phys. Rev. E},
  volume = {79},
  issue = {1},
  pages = {011111},
  numpages = {12},
  year = {2009},
  month = {Jan},
  publisher = {American Physical Society},
  doi = {10.1103/PhysRevE.79.011111},
  url = {https://link.aps.org/doi/10.1103/PhysRevE.79.011111}
}

@article{Vojta2012,
  title = {Monte Carlo simulations of the clean and disordered contact process in three dimensions},
  author = {Vojta, Thomas},
  journal = {Phys. Rev. E},
  volume = {86},
  issue = {5},
  pages = {051137},
  numpages = {11},
  year = {2012},
  month = {Nov},
  publisher = {American Physical Society},
  doi = {10.1103/PhysRevE.86.051137},
  url = {https://link.aps.org/doi/10.1103/PhysRevE.86.051137}
}

@article{Eisler2016,
  title = {Entanglement negativity in two-dimensional free lattice models},
  author = {Eisler, Viktor and Zimbor\'as, Zolt\'an},
  journal = {Phys. Rev. B},
  volume = {93},
  issue = {11},
  pages = {115148},
  numpages = {10},
  year = {2016},
  month = {Mar},
  publisher = {American Physical Society},
  doi = {10.1103/PhysRevB.93.115148},
  url = {https://link.aps.org/doi/10.1103/PhysRevB.93.115148}
}

@article{Bennett1996,
  title = {Concentrating partial entanglement by local operations},
  author = {Bennett, Charles H. and Bernstein, Herbert J. and Popescu, Sandu and Schumacher, Benjamin},
  journal = {Phys. Rev. A},
  volume = {53},
  issue = {4},
  pages = {2046--2052},
  numpages = {0},
  year = {1996},
  publisher = {American Physical Society},
  doi = {10.1103/PhysRevA.53.2046},
  url = {https://link.aps.org/doi/10.1103/PhysRevA.53.2046}
}

@article{Pandey2023,
	title = {Random geometry at an infinite-randomness fixed point},
	volume = {108},
	url = {https://link.aps.org/doi/10.1103/PhysRevB.108.064201},
	doi = {10.1103/PhysRevB.108.064201},
	number = {6},
	urldate = {2025-06-13},
	journal = {Phys. Rev. B},
	author = {Pandey, Akshat and Mahadevan, Aditya and Cowsik, Aditya},
	year = {2023},
	//note= {Publisher: American Physical Society},
	pages = {064201}
}

@article{kovacs2020-contactprocess,
	title = {Emergence of disconnected clusters in heterogeneous complex systems},
	volume = {10},
	copyright = {2020 The Author(s)},
	issn = {2045-2322},
	url = {https://www.nature.com/articles/s41598-020-78769-2},
	doi = {10.1038/s41598-020-78769-2},
	number = {1},
	urldate = {2025-04-16},
	journal = {Sci. Rep.},
	author = {Kovács, István A. and Juhász, Róbert},
	year = {2020},
	//note= {Publisher: Nature Publishing Group},
	pages = {1--8}
}

@article{Igloi2012-dynamics,
	title = {Entanglement entropy dynamics of disordered quantum spin chains},
	volume = {85},
	url = {https://link.aps.org/doi/10.1103/PhysRevB.85.094417},
	doi = {10.1103/PhysRevB.85.094417},
	number = {9},
	urldate = {2025-06-13},
	journal = {Phys. Rev. B},
	author = {Iglói, Ferenc and Szatmári, Zsolt and Lin, Yu-Cheng},
	year = {2012},
	//note= {Publisher: American Physical Society},
	pages = {094417}
}

@article{Kovacs2014-pottscorners,
	title = {Corner contribution to cluster numbers in the {Potts} model},
	volume = {89},
	doi = {10.1103/PhysRevB.89.064421},
	number = {6},
	urldate = {2025-06-03},
	journal = {Phys. Rev. B},
	author = {Kovács, István A. and Elçi, Eren Metin and Weigel, Martin and Iglói, Ferenc},
	year = {2014},
	//note= {Publisher: American Physical Society},
	pages = {064421}
}

@article{laflorencie2005,
	title = {Scaling of entanglement entropy in the random singlet phase},
	volume = {72},
	url = {https://link.aps.org/doi/10.1103/PhysRevB.72.140408},
	doi = {10.1103/PhysRevB.72.140408},
	number = {14},
	urldate = {2025-05-06},
	journal = {Phys. Rev. B},
	author = {Laflorencie, Nicolas},
	month = {Oct},
	year = {2005},
	//note= {Publisher: American Physical Society},
	pages = {140408}
}

@article{cardy1988,
	title = {Finite-size dependence of the free energy in two-dimensional critical systems},
	volume = {300},
	issn = {0550-3213},
	url = {https://www.sciencedirect.com/science/article/pii/0550321388906049},
	doi = {10.1016/0550-3213(88)90604-9},
	urldate = {2025-05-06},
	journal = {Nuclear Physics B},
	author = {Cardy, John L. and Peschel, Ingo},
	month = jan,
	year = {1988},
	pages = {377--392}
}

@article{vidal2003,
	title = {Entanglement in Quantum Critical Phenomena},
	volume = {90},
	url = {https://link.aps.org/doi/10.1103/PhysRevLett.90.227902},
	doi = {10.1103/PhysRevLett.90.227902},
	number = {22},
	urldate = {2025-05-06},
	journal = {Phys. Rev. Lett.},
	author = {Vidal, G. and Latorre, J. I. and Rico, E. and Kitaev, A.},
	month = {Jun},
	year = {2003},
	pages = {227902}
}

@article{holzhey1994,
	title = {Geometric and renormalized entropy in conformal field theory},
	volume = {424},
	issn = {0550-3213},
	url = {http://www.scopus.com/inward/record.url?scp=2442674839&partnerID=8YFLogxK},
	doi = {10.1016/0550-3213(94)90402-2},
	number = {3},
	urldate = {2025-05-06},
	journal = {Nuclear Physics, Section B},
	author = {Holzhey, Christoph and Larsen, Finn and Wilczek, Frank},
	month = {Aug},
	year = {1994},
	pages = {443--467}
}

@article{Helmes2016,
  title = {Universal corner entanglement of {D}irac fermions and gapless bosons from the continuum to the lattice},
  author = {Helmes, Johannes and Hayward Sierens, Lauren E. and Chandran, Anushya and Witczak-Krempa, William and Melko, Roger G.},
  journal = {Phys. Rev. B},
  volume = {94},
  issue = {12},
  pages = {125142},
  numpages = {14},
  year = {2016},
  month = {Sep},
  publisher = {American Physical Society},
  doi = {10.1103/PhysRevB.94.125142},
  url = {https://link.aps.org/doi/10.1103/PhysRevB.94.125142}
}

@article{calabrese2004,
	title = {Entanglement entropy and quantum field theory},
	volume = {2004},
	issn = {1742-5468},
	url = {https://dx.doi.org/10.1088/1742-5468/2004/06/P06002},
	doi = {10.1088/1742-5468/2004/06/P06002},
	number = {06},
	urldate = {2025-04-29},
	journal = {J. Stat. Mech.},
	author = {Calabrese, Pasquale and Cardy, John},
	month = {Jun},
	year = {2004},
	pages = {P06002}
}

@article{faulkner2016,
	title = {Shape dependence of entanglement entropy in conformal field theories},
	volume = {2016},
	issn = {1029-8479},
	url = {https://doi.org/10.1007/JHEP04(2016)088},
	doi = {10.1007/JHEP04(2016)088},
	number = {4},
	urldate = {2025-04-29},
	journal = {J. High Energ. Phys.},
	author = {Faulkner, Thomas and Leigh, Robert G. and Parrikar, Onkar},
	month = {Apr},
	year = {2016},
	pages = {88}
}

@article{WK-corner-entanglement,
  title = {Universality of Corner Entanglement in Conformal Field Theories},
  author = {Bueno, Pablo and Myers, Robert C. and Witczak-Krempa, William},
  journal = {Phys. Rev. Lett.},
  volume = {115},
  issue = {2},
  pages = {021602},
  numpages = {6},
  year = {2015},
  month = {Jul},
  publisher = {American Physical Society},
  doi = {10.1103/PhysRevLett.115.021602},
  url = {https://link.aps.org/doi/10.1103/PhysRevLett.115.021602}
}

@article{WK-skeletal-entanglement,
  title = {Entanglement of Skeletal Regions},
  author = {Berthiere, Cl\'ement and Witczak-Krempa, William},
  journal = {Phys. Rev. Lett.},
  volume = {128},
  issue = {24},
  pages = {240502},
  numpages = {6},
  year = {2022},
  month = {Jun},
  publisher = {American Physical Society},
  doi = {10.1103/PhysRevLett.128.240502},
  url = {https://link.aps.org/doi/10.1103/PhysRevLett.128.240502}
}

@article{Helen-brain,
	title = {Unveiling universal aspects of the cellular anatomy of the brain},
	volume = {7},
	copyright = {2024 The Author(s)},
	issn = {2399-3650},
	url = {https://www.nature.com/articles/s42005-024-01665-y},
	doi = {10.1038/s42005-024-01665-y},
	number = {1},
	urldate = {2025-04-15},
	journal = {Commun. Phys.},
	author = {Ansell, Helen S. and Kovács, István A.},
	month = jun,
	year = {2024},
	//note= {Publisher: Nature Publishing Group},
	keywords = {Biological physics, Phase transitions and critical phenomena},
	pages = {1--11}
}

@article{Helen-percolation,
  title = {Cluster tomography in percolation},
  author = {Ansell, Helen S. and Frank, Samuel J. and Kov\'acs, Istv\'an A.},
  journal = {Phys. Rev. Res.},
  volume = {5},
  issue = {4},
  pages = {043218},
  numpages = {8},
  year = {2023},
  month = {Dec},
  publisher = {American Physical Society},
  doi = {10.1103/PhysRevResearch.5.043218},
  url = {https://link.aps.org/doi/10.1103/PhysRevResearch.5.043218}
}

@article{WK_entanglement_susceptibility,
  title = {Entanglement susceptibilities and universal geometric entanglement entropy},
  author = {Witczak-Krempa, William},
  journal = {Phys. Rev. B},
  volume = {99},
  issue = {7},
  pages = {075138},
  numpages = {7},
  year = {2019},
  month = {Feb},
  publisher = {American Physical Society},
  doi = {10.1103/PhysRevB.99.075138},
  url = {https://link.aps.org/doi/10.1103/PhysRevB.99.075138}
}

@article{Refael2009,
	title = {Criticality and entanglement in random quantum systems},
	volume = {42},
	issn = {1751-8121},
	url = {https://dx.doi.org/10.1088/1751-8113/42/50/504010},
	doi = {10.1088/1751-8113/42/50/504010},
	number = {50},
	urldate = {2025-01-10},
	journal = {J. Phys. A: Math. Theor.},
	author = {Refael, G and Moore, J E},
	month = dec,
	year = {2009},
	pages = {504010}
}

@article{Zaletel2011,
  title = {Logarithmic Terms in Entanglement Entropies of 2D Quantum Critical Points and {S}hannon Entropies of Spin Chains},
  author = {Zaletel, Michael P. and Bardarson, Jens H. and Moore, Joel E.},
  journal = {Phys. Rev. Lett.},
  volume = {107},
  issue = {2},
  pages = {020402},
  numpages = {4},
  year = {2011},
  month = {Jul},
  publisher = {American Physical Society},
  doi = {10.1103/PhysRevLett.107.020402},
  url = {https://link.aps.org/doi/10.1103/PhysRevLett.107.020402}
}

@article{Rokhsar1988,
  title = {Superconductivity and the Quantum Hard-Core Dimer Gas},
  author = {Rokhsar, Daniel S. and Kivelson, Steven A.},
  journal = {Phys. Rev. Lett.},
  volume = {61},
  issue = {20},
  pages = {2376--2379},
  numpages = {0},
  year = {1988},
  month = {Nov},
  publisher = {American Physical Society},
  doi = {10.1103/PhysRevLett.61.2376},
  url = {https://link.aps.org/doi/10.1103/PhysRevLett.61.2376}
}

@article{Fradkin2006,
  title = {Entanglement Entropy of 2D Conformal Quantum Critical Points: Hearing the Shape of a Quantum Drum},
  author = {Fradkin, Eduardo and Moore, Joel E.},
  journal = {Phys. Rev. Lett.},
  volume = {97},
  issue = {5},
  pages = {050404},
  numpages = {4},
  year = {2006},
  month = {Aug},
  publisher = {American Physical Society},
  doi = {10.1103/PhysRevLett.97.050404},
  url = {https://link.aps.org/doi/10.1103/PhysRevLett.97.050404}
}

@article{Fisher1999,
	title = {{Phase transitions and singularities in random quantum systems}},
	volume = {263},
	issn = {03784371},
	doi = {10.1016/S0378-4371(98)00498-1},
	journal = {Physica A: Statistical Mechanics and its Applications},
	author = {Fisher, Daniel S.},
	year = {1999},
	pages = {222--233}
}

@article{Zou2022-multipartite,
  title = {Multipartite entanglement in the random {I}sing chain},
  author = {Zou, Jay S. and Ansell, Helen S. and Kov\'acs, Istv\'an A.},
  journal = {Phys. Rev. B},
  volume = {106},
  issue = {5},
  pages = {054201},
  numpages = {7},
  year = {2022},
  month = {Aug},
  publisher = {American Physical Society},
  doi = {10.1103/PhysRevB.106.054201},
  url = {https://link.aps.org/doi/10.1103/PhysRevB.106.054201}
}

@article{Kovacs20143dperc,
  title = {Corner contribution to percolation cluster numbers in three dimensions},
  author = {Kov\'acs, Istv\'an A. and Igl\'oi, Ferenc},
  journal = {Phys. Rev. B},
  volume = {89},
  issue = {17},
  pages = {174202},
  numpages = {5},
  year = {2014},
  month = {May},
  publisher = {American Physical Society},
  doi = {10.1103/PhysRevB.89.174202},
  url = {https://link.aps.org/doi/10.1103/PhysRevB.89.174202}
}

@article{Tagliacozzo2009,
  title = {Simulation of two-dimensional quantum systems using a tree tensor network that exploits the entropic area law},
  author = {Tagliacozzo, L. and Evenbly, G. and Vidal, G.},
  journal = {Phys. Rev. B},
  volume = {80},
  issue = {23},
  pages = {235127},
  numpages = {17},
  year = {2009},
  month = {Dec},
  publisher = {American Physical Society},
  doi = {10.1103/PhysRevB.80.235127},
  url = {https://link.aps.org/doi/10.1103/PhysRevB.80.235127}
}

@article{Song2011,
  title = {Entanglement entropy of the two-dimensional {H}eisenberg antiferromagnet},
  author = {Song, H. Francis and Laflorencie, Nicolas and Rachel, Stephan and Le Hur, Karyn},
  journal = {Phys. Rev. B},
  volume = {83},
  issue = {22},
  pages = {224410},
  numpages = {7},
  year = {2011},
  month = {Jun},
  publisher = {American Physical Society},
  doi = {10.1103/PhysRevB.83.224410},
  url = {https://link.aps.org/doi/10.1103/PhysRevB.83.224410}
}

@article{Kallin2011,
  title = {Anomalies in the entanglement properties of the square-lattice {H}eisenberg model},
  author = {Kallin, Ann B. and Hastings, Matthew B. and Melko, Roger G. and Singh, Rajiv R. P.},
  journal = {Phys. Rev. B},
  volume = {84},
  issue = {16},
  pages = {165134},
  numpages = {11},
  year = {2011},
  month = {Oct},
  publisher = {American Physical Society},
  doi = {10.1103/PhysRevB.84.165134},
  url = {https://link.aps.org/doi/10.1103/PhysRevB.84.165134}
}

@article{Eisert2010,
  title = {{Colloquium: Area laws for the entanglement entropy}},
  author = {Eisert, J. and Cramer, M. and Plenio, M. B.},
  journal = {Rev. Mod. Phys.},
  volume = {82},
  issue = {1},
  pages = {277--306},
  numpages = {0},
  year = {2010},
  publisher = {American Physical Society},
  doi = {10.1103/RevModPhys.82.277},
  url = {https://link.aps.org/doi/10.1103/RevModPhys.82.277}
}

@article{Calabrese2009,
doi = {10.1088/1751-8121/42/50/500301},
url = {https://dx.doi.org/10.1088/1751-8121/42/50/500301},
year = {2009},
publisher = {},
volume = {42},
number = {50},
pages = {500301},
author = {Pasquale Calabrese and John Cardy and Benjamin Doyon},
title = {{Entanglement entropy in extended quantum systems}},
journal = {Journal of Physics A: Mathematical and Theoretical}
}

@article{Friedemann2018,
    author = {Friedemann, Sven and Duncan, Will J. and Hirschberger, Max and Bauer, Thomas W. and K\"uchler, Robert and Neubauer, Andreas and Brando Manuel and Fleiderer, Christian and Grosche, F. Malte},
    title = {{Quantum tricritical points in NbFe$_2$}},
    journal = {Nature Physics},
    year = {2018},
    pages = {62-67},
    volume = {14},
    number = {1},
    doi = {10.1038/nphys4242},
    url = {https://doi.org/10.1038/nphys4242}
}

@article{Bitko1996,
  title = {Quantum Critical Behavior for a Model Magnet},
  author = {Bitko, D. and Rosenbaum, T. F. and Aeppli, G.},
  journal = {Phys. Rev. Lett.},
  volume = {77},
  issue = {5},
  pages = {940--943},
  numpages = {0},
  year = {1996},
  publisher = {American Physical Society},
  doi = {10.1103/PhysRevLett.77.940},
  url = {https://link.aps.org/doi/10.1103/PhysRevLett.77.940}
}

@article{Silevitch2010,
author = {D. M. Silevitch  and G. Aeppli  and T. F. Rosenbaum },
title = {Switchable hardening of a ferromagnet at fixed temperature},
journal = {Proceedings of the National Academy of Sciences},
volume = {107},
number = {7},
pages = {2797-2800},
year = {2010},
doi = {10.1073/pnas.0910575107},
URL = {https://www.pnas.org/doi/abs/10.1073/pnas.0910575107},
eprint = {https://www.pnas.org/doi/pdf/10.1073/pnas.0910575107}}

@article{Kovacs2012-percolation,
   title={Corner contribution to percolation cluster numbers},
   author = {Kov\'acs, Istv\'an A. and Igl\'oi, Ferenc and Cardy, John},
  journal = {Phys. Rev. B},
  volume = {86},
  issue = {21},
  pages = {214203},
  numpages = {6},
  year = {2012},
  publisher = {American Physical Society},
  doi = {10.1103/PhysRevB.86.214203},
  url = {https://link.aps.org/doi/10.1103/PhysRevB.86.214203}
}

@article{Kovacs2012-rtfim,
   title={{Universal logarithmic terms in the entanglement entropy of 2d, 3d and 4d random transverse-field Ising models}},
   volume={97},
   DOI={10.1209/0295-5075/97/67009},
   number={6},
   journal={EPL (Europhysics Letters)},
   publisher={IOP Publishing},
   author={Kovács, I. A. and Iglói, F.},
   year={2012},
   pages={67009} }

@article{Kovacs2010-2d-sdrg,
   title={{Renormalization group study of the two-dimensional random transverse-field Ising model}},
   author = {Kov\'acs, Istv\'an A. and Igl\'oi, Ferenc},
  journal = {Phys. Rev. B},
  volume = {82},
  issue = {5},
  pages = {054437},
  numpages = {13},
  year = {2010},
  publisher = {American Physical Society},
  doi = {10.1103/PhysRevB.82.054437},
  url = {https://link.aps.org/doi/10.1103/PhysRevB.82.054437}
}

@article{Kovacs2009-rtfim,
  title = {{Critical behavior and entanglement of the random transverse-field Ising model between one and two dimensions}},
  author = {Kov\'acs, Istv\'an A. and Igl\'oi, Ferenc},
  journal = {Phys. Rev. B},
  volume = {80},
  issue = {21},
  pages = {214416},
  numpages = {9},
  year = {2009},
  publisher = {American Physical Society},
  doi = {10.1103/PhysRevB.80.214416},
  url = {https://link.aps.org/doi/10.1103/PhysRevB.80.214416}
}

@article{Kovacs2011-sdrg,
doi = {10.1088/0953-8984/23/40/404204},
url = {https://dx.doi.org/10.1088/0953-8984/23/40/404204},
year = {2011},
publisher = {},
volume = {23},
number = {40},
pages = {404204},
author = {István A. Kovács and Ferenc Iglói},
title = {Renormalization group study of random quantum magnets},
journal = {Journal of Physics: Condensed Matter}
}

@article{Kovacs2011-3d-sdrg,
  title = {Infinite-disorder scaling of random quantum magnets in three and higher dimensions},
  author = {Kov\'acs, Istv\'an A. and Igl\'oi, Ferenc},
  journal = {Phys. Rev. B},
  volume = {83},
  issue = {17},
  pages = {174207},
  numpages = {5},
  year = {2011},
  publisher = {American Physical Society},
  doi = {10.1103/PhysRevB.83.174207},
  url = {https://link.aps.org/doi/10.1103/PhysRevB.83.174207}
}

@article{Kovacs2022-qmcp,
  title = {{Quantum multicritical point in the two- and three-dimensional random transverse-field Ising model}},
  author = {Kov\'acs, Istv\'an A.},
  journal = {Phys. Rev. Res.},
  volume = {4},
  issue = {1},
  pages = {013072},
  numpages = {8},
  year = {2022},
  publisher = {American Physical Society},
  doi = {10.1103/PhysRevResearch.4.013072},
  url = {https://link.aps.org/doi/10.1103/PhysRevResearch.4.013072}
}

@article{Kovacs2024-qmcp,
  title = {{Quantum entanglement in the multicritical disordered Ising model}},
  author = {Kov\'acs, Istv\'an A.},
  journal = {Phys. Rev. B},
  volume = {109},
  issue = {21},
  pages = {214202},
  numpages = {6},
  year = {2024},
  publisher = {American Physical Society},
  doi = {10.1103/PhysRevB.109.214202},
  url = {https://link.aps.org/doi/10.1103/PhysRevB.109.214202}
}

@article{Ma1976,
  title = {Random Antiferromagnetic Chain},
  author = {Ma, Shang-keng and Dasgupta, Chandan and Hu, Chin-kun},
  journal = {Phys. Rev. Lett.},
  volume = {43},
  issue = {19},
  pages = {1434--1437},
  numpages = {0},
  year = {1979},
  publisher = {American Physical Society},
  doi = {10.1103/PhysRevLett.43.1434},
  url = {https://link.aps.org/doi/10.1103/PhysRevLett.43.1434}
}

@article{Dasgupta1980,
  title = {{Low-temperature properties of the random {H}eisenberg antiferromagnetic chain}},
  author = {Dasgupta, Chandan and Ma, Shang-keng},
  journal = {Phys. Rev. B},
  volume = {22},
  issue = {3},
  pages = {1305--1319},
  numpages = {0},
  year = {1980},
  publisher = {American Physical Society},
  doi = {10.1103/PhysRevB.22.1305},
  url = {https://link.aps.org/doi/10.1103/PhysRevB.22.1305}
}

\section*{Appendix A: Geometric Method}\label{appedixA}

With PBC, the corner contribution can be ascertained precisely by employing a geometric approach which compares the entanglement entropy of different subsystem geometries to isolate the corners. This method is applicable to both line segments \cite{Helen-percolation} and sheared squares \cite{Kovacs2012-rtfim}. A sheared square means a subsystem whose linear extent is half the linear size of the system, $\ell=L/2$, and whose (smaller) interior angle satisfies $\gamma=\tan^{-1}(1/n)$ for $n=0,1,2,\dots$ (figure \ref{fig:shear-square}). 

\begin{figure}[hbt]
    \centering
    \includegraphics[width=0.8\linewidth]{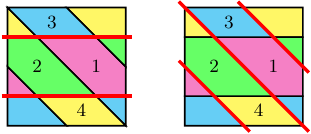}
        \vskip -0.2cm
    \caption{\textbf{The geometric method illustrated for a sheared square with $n=1$.} The system is partitioned into 4 equal subsystems $\{1,2,3,4\}$, which can be combined into 4 slabs $\{(1,2),(3,4)\}$ (left) and $\{(1,3),(2,4)\}$ (right). The combined boundary of the slabs (red) is equal to the combined boundary of the squares, although without corners.}
    \label{fig:geom-method}
\end{figure}

For such geometries, the system can be partitioned into 4 disjoint (sheared) square subsystems of equal area (figure \ref{fig:geom-method}). With this partitioning, we can also form 4 slab subsystems by joining adjacent square subsystems. In practice, the ``slanted'' boundary is defined on a finite square lattice as $1/n$ step, which ensures that PBC are satisfied. 
The accumulated boundary of all the square subsystems is exactly equal to the accumulated boundary of the slab subsystems; however due to PBC the slab geometry lacks only the corners where the four square subsystems meet. Therefore, any contribution coming from the boundary is the same for the two geometries, canceling out the area law in each sample for each subsystem position, together with finite-size corrections and statistical fluctuations.
Yet, the entanglement entropy of the square geometry $\Scal_{sq}$ is not necessarily equal to the entanglement entropy of the slab geometry $\Scal_{sl}$, as there may be ``corner'' clusters which contribute to the entanglement entropy of all four slabs, but not all four squares. These corner clusters are responsible for the corner contribution, which is directly obtained as the difference between the cluster count in the two geometries \cite{Kovacs2012-rtfim}.

\section*{Appendix B: Cluster Counting Procedure}

The entanglement entropy of a subsystem was calculated by first counting the number of sites of each cluster contained within the subsystem. If, for a given cluster, the number of cluster sites inside the subsystem is less than the total mass of the cluster, that cluster contributes 1 to the entanglement entropy; otherwise it does not contribute. For the first iteration of the procedure, the entire sample must be traversed once to obtain cluster counts for the 4 (sheared) square subsystems defined above for the geometric method. The slab cluster numbers can then be easily obtained by adding the cluster counts of its two constituent (sheared) squares, and the difference of the two values of entanglement entropy gives the corner contribution. This basic protocol is repeated for every possible placement of the (sheared) square subsystem within the sample. However, for subsequent iterations, where the subsystem $1$, say, is moved exactly one unit to the right of where it was located in the previous calculation, the number of sites of each cluster will be largely the same. These counts can be updated by just looking at the two boundary layers of $L$ sites which are not in the intersection of the two subsystems. This way, calculating the average corner contribution for all subsystem location in a sample takes $\mathcal{O}(L^3)$ time. In practice, calculating the average corner contribution for a critical sample of linear size $L=128$ with box-$h$ disorder, takes approximately 5 CPU seconds.

\end{document}